%% file: main.tex
\documentclass[conference]{IEEEtran}

\include{macros}

\begin{document}

\title{An Empirical Analysis of Compatibility Issues for \\ Industrial Mobile Games}


\author{
    \centering
    \IEEEauthorblockN{Zihe Song}
    \IEEEauthorblockA{
    \textit{University of Texas at Dallas}\\
    Richardson, TX, USA \\
    zihe.song@utdallas.edu}
    \and
    \IEEEauthorblockN{Yingfeng Chen}
    \IEEEauthorblockA{
    \textit{NetEase Fuxi AI Lab}\\
    Hangzhou, China \\
    chenyingfeng1@corp.netease.com}
    \and
    \IEEEauthorblockN{Lei Ma}
    \IEEEauthorblockA{
    \textit{Edmonton, Alberta, Canada}\\
    Edmonton, Alberta, Canada \\
    ma.lei@acm.org}
    \and
    \IEEEauthorblockN{Shangjie Lu}
    \IEEEauthorblockA{
    \textit{NetEase Inc.}\\
    Hangzhou, China \\
    lushangjie@corp.netease.com}

    \and
    \IEEEauthorblockN{Honglei Lin}
    \IEEEauthorblockA{
    \textit{NetEase Inc.}\\
    Hangzhou, China \\
    hzlinhonglei3@corp.netease.com}
    
    \and
    \IEEEauthorblockN{Changjie Fan}
    \IEEEauthorblockA{
    \textit{NetEase Fuxi AI Lab}\\
    Hangzhou, China \\
    fanchangjie@corp.netease.com}
    
    \and
    \IEEEauthorblockN{Wei Yang}
    \IEEEauthorblockA{
    \textit{University of Texas at Dallas}\\
    Richardson, TX, USA \\
    wei.yang@utdallas.edu}
}

\maketitle

\begin{abstract}
Detecting and fixing compatibility issues become increasingly important for mobile game development. The constant evolution of mobile operating systems and the severe fragmentation of mobile devices makes it challenging for game developers to detect and fix compatibility issues in time for various device models. The undetected compatibility issues can ruin the user experience, and cause financial loss to game companies and players. Unfortunately, up to the present, mobile game testing is still rather challenging in general. The pressing compatibility issue of mobile games is largely untouched in the research community so far.

To bridge the gap, in this experience paper, we perform an empirical study on common compatibility issues of popular commercial mobile games. In particular, we select four active and representative mobile games with well-documented bug reports, containing over seven million lines of code and over 20,000 commits over the past several years. We successfully create a dataset with complete information about bugs and bug fixing details, to investigate the common compatibility issues and fixing strategies. We performed an in-depth manual inspection of the most common symptoms and root causes of these compatibility issues, and analyzed the common fixing strategies of issues under each root cause category. We believe our findings and implications are useful for developers in addressing compatibility hurdles during the developing process. Our results also provide insights for future research on compatibility issue testing and bug fixing for mobile games.
\end{abstract}


\section{Introduction}
\label{sec:intro}
\input{introduction.tex}

\section{Background}
\label{sec:practice}
\input{testingpractice.tex}

\section{Methodology}
\label{sec:method}
\input{methodology.tex}

\section{RQ1: Symptom Analysis}
\label{sec:manifestation}
\input{manifestation.tex}

\section{RQ2: Root Cause Analysis}
\label{sec:rootcause}
\input{rootcause.tex}

\section{RQ3: Issue Fixing}
\label{sec:fixing}
\input{fixing.tex}

\section{Threats to Validity}
\label{sec:validity}
\input{validity.tex}

\section{Related work}
\label{sec:related}
\input{related.tex}

\section{Conclusion}
\label{sec:conclusion}
\input{conclusion.tex}

\section*{Acknowledgment}
This work was partially supported by Siemens Fellowship and NSF grant CCF-2146443. We thank our industrial research partner \netease, especially the Fuxi AI Lab for their discussion and support with the experiments. 

\bibliographystyle{IEEEtran}
\bibliography{isstareference}

\end{document}

%% file: macros.tex
\usepackage{xspace}    
\usepackage{booktabs}
\usepackage{color}
\usepackage{xcolor}
\usepackage{makecell,rotating}
\usepackage{multirow}
\usepackage{multicol}
\usepackage{colortbl}
\usepackage{float}
\usepackage{placeins}
\usepackage{tikz}
\usepackage{subfigure}
\usepackage{algorithm}
\usepackage{algorithmicx}
\usepackage{algpseudocode}
\usepackage{hyperref}

\usepackage{amsmath,amsfonts}
\usepackage{graphicx}
\usepackage{textcomp}
\usepackage{xcolor}
\usepackage{listings}

\lstdefinestyle{customcsharp}{
  language=[Sharp]C,
  basicstyle=\ttfamily\footnotesize,
  numbers=left,
  numberstyle=\tiny,
  stepnumber=1,
  numbersep=8pt,
  backgroundcolor=\color{white},
  showspaces=false,
  showstringspaces=false,
  breaklines=true,
  frame=single,
  keywordstyle=\color{blue},
  commentstyle=\color{gray},
  stringstyle=\color{red}
}
\lstdefinelanguage{json}{
  basicstyle=\ttfamily\small,
  numbers=left,
  numberstyle=\tiny,
  stepnumber=1,
  numbersep=8pt,
  showstringspaces=false,
  breaklines=true,
  frame=single,
  backgroundcolor=\color{white},
  literate=
    *{0}{{{\color{black}0}}}{1}
     {1}{{{\color{black}1}}}{1}
     {2}{{{\color{black}2}}}{1}
     {3}{{{\color{black}3}}}{1}
     {4}{{{\color{black}4}}}{1}
     {5}{{{\color{black}5}}}{1}
     {6}{{{\color{black}6}}}{1}
     {7}{{{\color{black}7}}}{1}
     {8}{{{\color{black}8}}}{1}
     {9}{{{\color{black}9}}}{1}
     {:}{{{\color{black}:}}}{1}
     {,}{{{\color{black},}}}{1}
     {"}{{{\color{red}"}}}{1},
  keywordstyle=\color{green},
  morekeywords={true,false,null}
}
\usepackage{enumitem}
\usepackage{tcolorbox}
\usepackage{longtable}
\usepackage{tabu}
\usepackage{balance}

\newcommand{\eg}{{\it e.g.,}\xspace}
\newcommand{\etal}{{\it et al.}\xspace}
\newcommand{\etc}{{\it etc.}\xspace}
\newcommand{\ie}{{\it i.e.,}\xspace}

\definecolor{comment-red}{rgb}{0.8,0,0}


\def\netease{NetEase, Inc.\xspace}

%% file: introduction.tex

With the leap of mobile hardware performance in the past decades, mobile game industry has achieved huge momentum. More and more computational-intensive games that were  provided only on PC/laptops before, are currently well supported by mobile devices. According to Newzoo's 2020 Global Games Market Report \cite{newzoo}, mobile game is the largest segment by far among all gaming platforms, \textit{i.e.} mobile game revenues grew 13.3\% year on year and occupied 49\% of the global games market. The booming market makes the quality assurance for mobile game software of great importance. 
However, quality assurance of mobile video games is rather challenging, which requires heavy user interaction and a certain level of intelligence~\cite{AleemCA16,wuji, 9240641}. Even to this moment, the major industrial mobile game producers still heavily rely on human testers for game testing~\cite{baek2016}.

Compatibility issue is one of the most pressing quality problems and pain points in mobile game quality assurance. 
Generally, mobile compatibility issues refer to the problems that the app cannot run as expected on some specific operating systems, hardware, or devices. 
Compatibility issues on general mobile applications were well-studied in recent years \cite{wei2016,han12,ham2014}, many of which are mainly caused by severe \emph{system fragmentation}. 

Currently, there are five versions of Apple systems (include iOS and iPadOS) and nine versions of Android systems still in concurrent usage. API specifications and development guidelines are constantly changing with the system evolution.
Moreover, mobile manufacturers choose to equip Android systems on their models with distinct customization, these modified versions may contain bugs, or do not fully comply with the original Android specifications. For software developers, designing and testing their apps to adapt to all kinds of APIs and model combinations is a huge and even impractical challenge. According to study in 2020~\cite{guilardi2020}, only 60\% of apps could be adapted to the newly released Android version in a month. Most apps could be adapted after 12 months of release.

However, the compatibility analysis of general mobile software is not fully applicable to mobile games\cite{ki2019,nighthawk}. The complicated user interface (UI) layouts and artistic styles make mobile games prone to two other types of compatibility issues caused by hardware, \emph{screen shape adaption} and \emph{computation units adaption}. 
Mobile manufacturers prefer to customize their screen shapes to create a full-screen immersion. Different resolutions and front camera positions are prone to cause compatibility issues such as UI blocking or element dislocation. Also, exquisite game graphics relies on rendering techniques, the performance bottlenecks in computation units on specific hardware or low-configuration models may bring compatibility issues.

Game compatibility issues caused by \emph{System fragmentation} and \emph{hardware adaptation} occur frequently, taking developers' tons of time and manual efforts on compatibility testing. An effective automated compatibility testing tool for mobile games is in urgent demand.
To make the first step in automated compatibility testing of mobile games, we conduct an in-depth empirical study to investigate the common compatibility issues of mobile games. We collect and analyze compatibility issues from four representative industrial mobile games. These games are of different types, currently still operating with a large number of daily activeusers, and containing more than seven million lines of code and 20,000 commits in total. With these collected industrial mobile game codebases, we mainly investigate the following research questions:

\begin{itemize}[leftmargin=*]

\item \textbf{RQ1: What are the most common symptoms of mobile game compatibility issues?} Finding the common symptoms of compatibility issues can facilitate the design and development of better automated detection techniques.

\item \textbf{RQ2: What are the most common root causes of mobile game compatibility issues?} 
Analyzing the root causes of compatibility issues can help game developers avoid creating compatibility issues during development phase, also benefit the further automated debugging techniques.

\item \textbf{RQ3: What are the most common fixing strategies for mobile game compatibility issues?} By summarizing common fixing strategies of compatibility issues, we provide some insights that hope to help mobile game developers fix related issues and guide future researchers to design automatically detecting and fixing techniques.
\end{itemize}

We hope the reported results of this experience paper could raise the attention of researchers and practitioners to propose more effective testing and analysis techniques towards addressing mobile game compatibility challenges.

%% file: testingpractice.tex
In this section, we introduce the real-world mobile game compatibility testing practice at \netease, which is one of the world-leading game companies.

Compatibility testing is an important indicator of mobile game quality.
Different from functional testing, compatibility issues affect only parts of devices and players in general. However, if an issue happens on a mainstream model that occupies a large market partition, the problem will escalate and potentially lead to severe consequences. Based on this consideration, compatibility testing become a general requirement for various types of mobile apps.

Compatibility testing is performed at some critical milestones, such as the pre-Beta or closed-Beta stage, rather than being implemented as a regular regression test suite (\ie compiled with batched commits by game developers). Because if an app passes the compatibility testing but fails the other testing procedures (\eg functional testing or weak network testing), then passing compatibility tests can be meaningless because fixing other testing issues may introduce new compatibility issues.

In general, there are three candidate methods for testing mobile games, physical models, emulators, and cloud platforms. 
\begin{itemize}[leftmargin=*]
\item \textbf{Physical models}: Using physical devices is a reasonable approach because it will be the real environment for game playing. However, the expenditure on the purchase and maintenance of the device pool could be expensive. 

\item \textbf{Emulators}: Using emulators for compatibility testing is an economical choice. But it is insufficient and incomplete to solely relying on emulators, because emulators cannot fully simulate real devices, resulting in missing potential issues that could have been observed on real models.

\item \textbf{Cloud platforms}: Cloud platforms can be a good candidate for compatibility testing. There exist several available business platforms for compatibility testing, but these platforms provide only the test results without testing methodologies and the incidents during testing. As the end-user, testers may not be aware of the detailed information once an error occurs. 
\end{itemize}

Considering that testing with physical models is fully transparent and the device pool can be shared by multiple development teams, \netease chose to maintain a mobile device pool with a wide range of models. Due to limited resources, not all models will be used while testing an app.
The testers will follow a series of processes to acquire the testing set of models. The decision is driven by the market circumstances, in particular,  selecting the most popular devices for each game based on the main user group of the game, which combines multiple factors such as model price range, players' survey information, and the model information of registered players, \etc

Currently, compatibility testing for all the games is performed by a single team in \netease. Most of the systematic non-random compatibility tests are performed \textit{manually} by the team members, some of which are also outsourcing to other specialized testing companies.
Certain routine procedures in compatibility testing are automated, \eg installation, uninstallation, and random testing.
Therefore, building a fully automated and intelligent testing infrastructure to enable systematic and synergistic compatibility testing is an urgent demand.

According to \netease's experience, mobile games have several unique challenges that require special attention.
First, the functional logic of mobile games is often complicated due to frequent interactions between players and apps, which leads to sophisticated testing procedures.
Because the available resource is limited, it is quite a challenge to incorporate compatibility testing into already complex game testing procedures.
Second, fixing compatibility issues is also challenging. It involves not only game source code, but also large amounts of multi-media resources such as graphic models and shaders.
As a result, mobile game compatibility issues cannot be hot patched (\ie patching the app without reinstalling). 



%% file: methodology.tex
The goal of this paper is to conduct a comprehensive study of the compatibility issues in real-world industrial mobile games, hoping to draw the attention of the research community to investigate and propose effective mobile compatibility testing techniques as the first essential step.

\subsection{Subject Mobile Game Selection}
To be representative, we collect relevant compatibility issue data from four real-world commercial mobile games released by industrial collaborator \netease, which covers different game types, having various amounts of code size and downloads, and still being active with different development time scopes (from months to years). All four games have been released, attracting more than two million regular players in total. 

\begin{figure}[htbp]
\centering
\subfigure[ACT Game BS]{
\includegraphics[width=3.5cm]{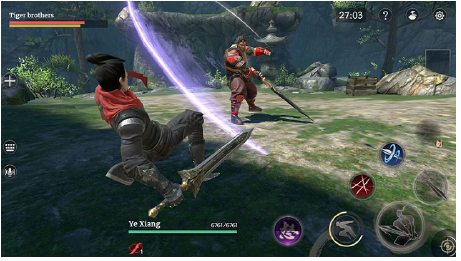}
}
\quad
\subfigure[RPG Game LJ]{
\includegraphics[width=3.5cm]{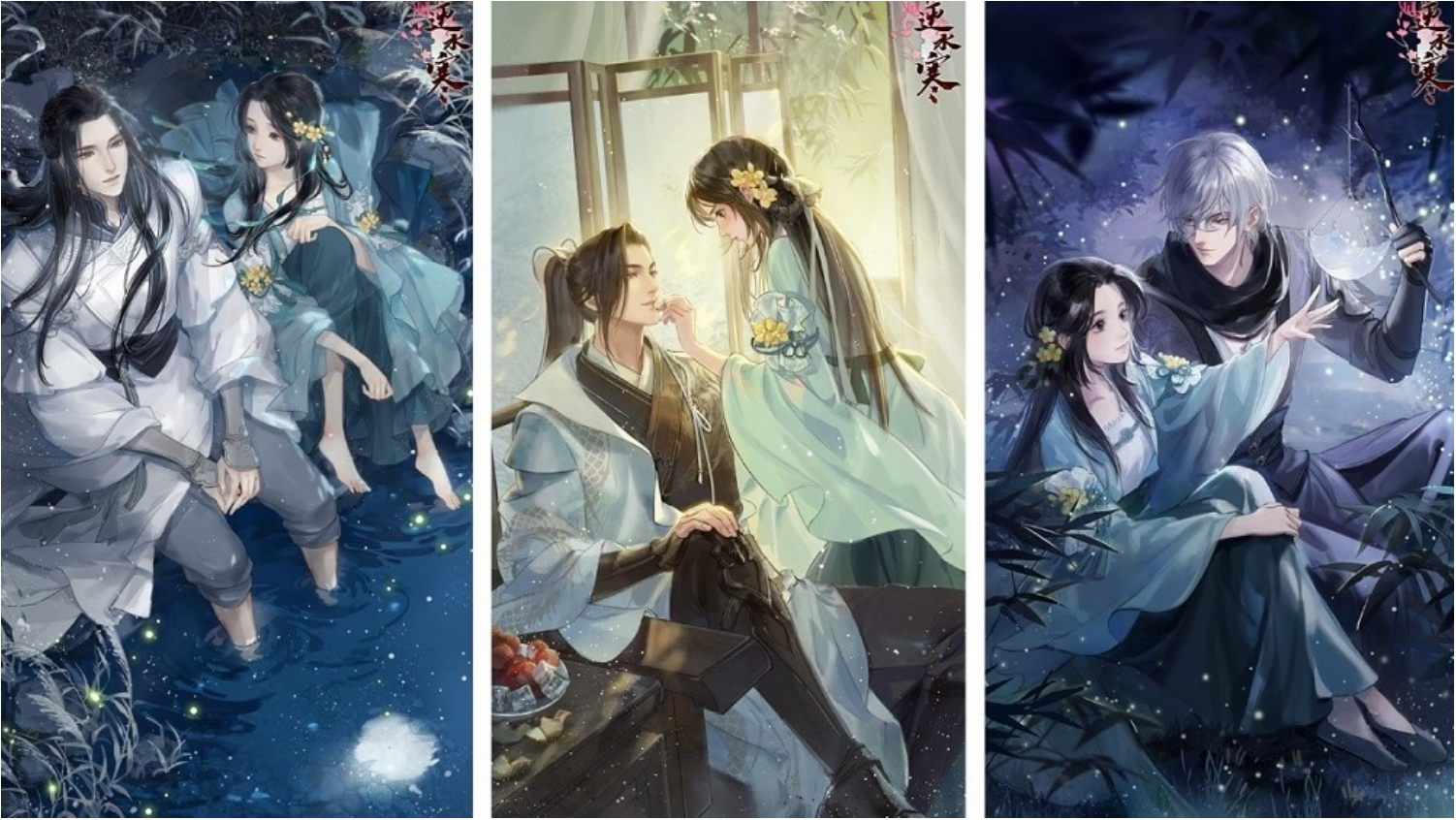}
}
\quad
\subfigure[BR Game EI]{
\includegraphics[width=3.5cm]{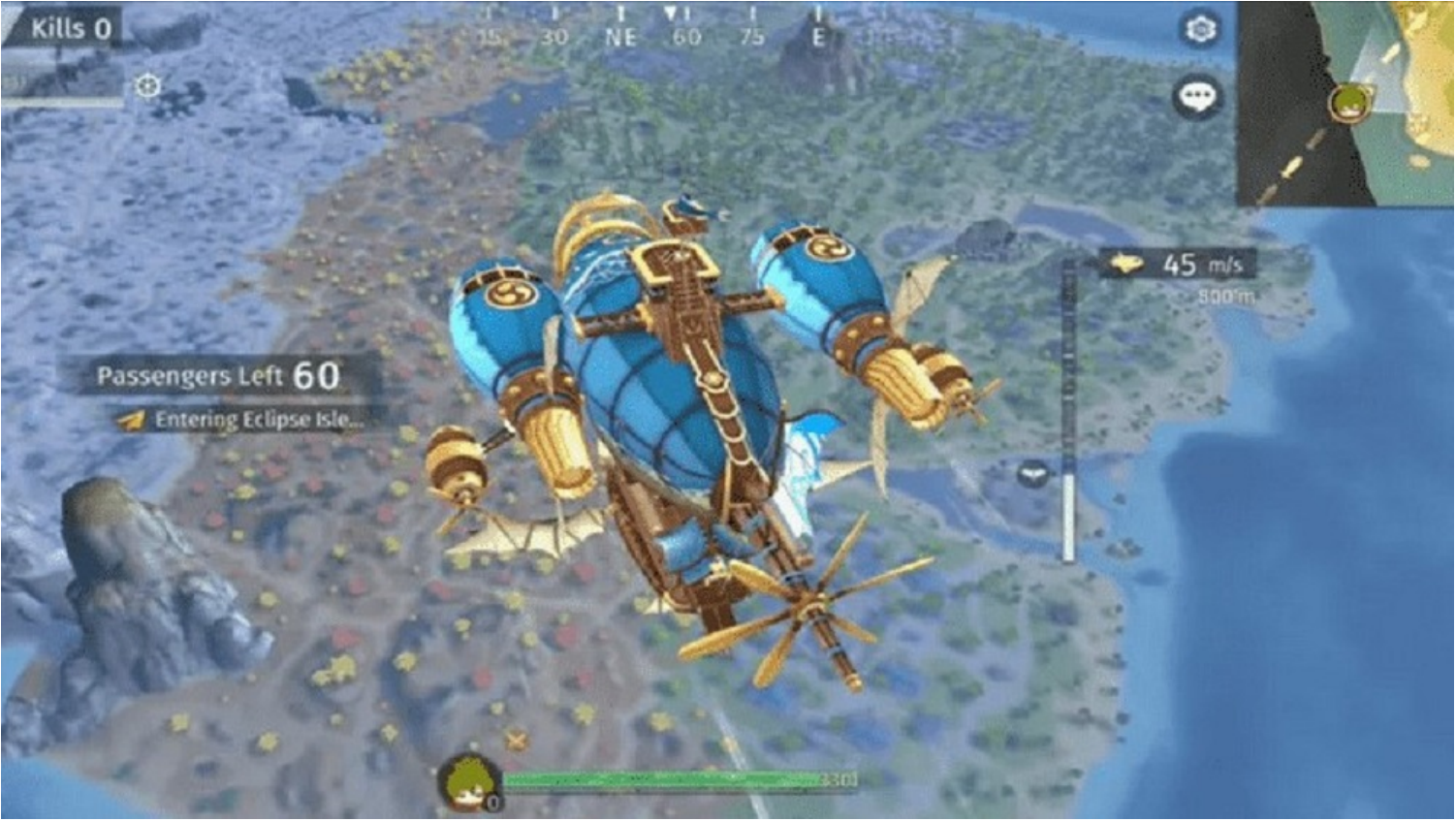}
}
\quad
\subfigure[MMORPG Game GS]{
\includegraphics[width=3.5cm]{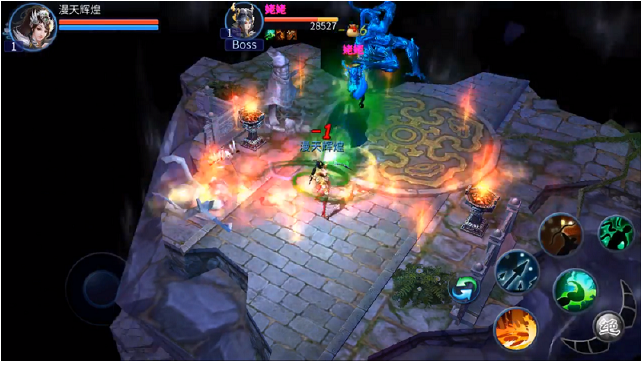}
}
\caption{Screenshots of the four investigated mobile games}
\vspace{-2mm}
\label{fig:games}
\end{figure}

Fig.~\ref{fig:games} shows the game screenshots and Table \ref{Tbl:gameinfo} summarizes the detailed information of these four mobile games. Overall, these four games contain more than seven million lines of code.

\begin{itemize}[leftmargin=*]
\item \textbf{BS} (Fig.~\ref{fig:games}~(a)) is an Instant Action game (ACT). The ACT game often adopts a large number of hitting effects, potentially causing GPU issues across mobile devices.

\item \textbf{LJ} (Fig.~\ref{fig:games}~(b)) is a lover Role Play Game (RPG). The RPG games are composed of a series of UI, commonly causing UI display problems.

\item \textbf{EI} (Fig.~\ref{fig:games}~(c)) is a Battle Royale game (BR), which is among the most prevalent type of games in recent years attracting millions of players.

\item \textbf{GS} (Fig.~\ref{fig:games}~(d)) is a Massively Multi-player Online Role-Playing Game (MMORPG), MMORPG games usually have a large open world for players to battle and socialize, and all kinds of compatibility issues may appear in MMORPG games. GS is a popular game and has been in service and maintenance for more than four years, providing a long timeline of bug reports for our analysis.
\end{itemize}

\begin{table*}[t]
\centering
\caption{A summary of the four investigated mobile games}
\vspace{-2mm}
\label{Tbl:gameinfo}
\begin{tabular}{c|c|c|c|c|c}
\toprule
\textbf{\begin{tabular}[c]{@{}c@{}}Game \\ Name\end{tabular}} & \textbf{\begin{tabular}[c]{@{}c@{}}Game\\ Type\end{tabular}} & \textbf{\begin{tabular}[c]{@{}c@{}}Lines Of\\ Code\end{tabular}} & \textbf{\begin{tabular}[c]{@{}c@{}}Development\\ Time\end{tabular}} & \textbf{\begin{tabular}[c]{@{}c@{}}Compatibility\\ Issues\end{tabular}} & \textbf{\begin{tabular}[c]{@{}c@{}}Bug Report In\\ Alpha Test Phase\end{tabular}} \\ \hline
BS                                                            & ACT                                                          & 1,470,000                                                        & 2017.6 $\sim$ Present                                               & 36                                                                      & 2,868        \\
LJ                                                            & RPG                                                          & 400,000                                                          & 2019.12 $\sim$ Present                                              & 85                                                                      & 2,102        \\
EI                                                            & BR                                                           & 810,000                                                          & 2017.5 $\sim$ Present                                               & 93                                                                      & 9,080     \\
GS                                                            & MMORPG                                                       & 4,600,000                                                        & 2015.4 $\sim$ Present                                               & 60                                                                      & 650                           \\ \bottomrule
\end{tabular}
\end{table*}

\begin{figure*}[t]
\centering
\includegraphics[width=0.7\linewidth]{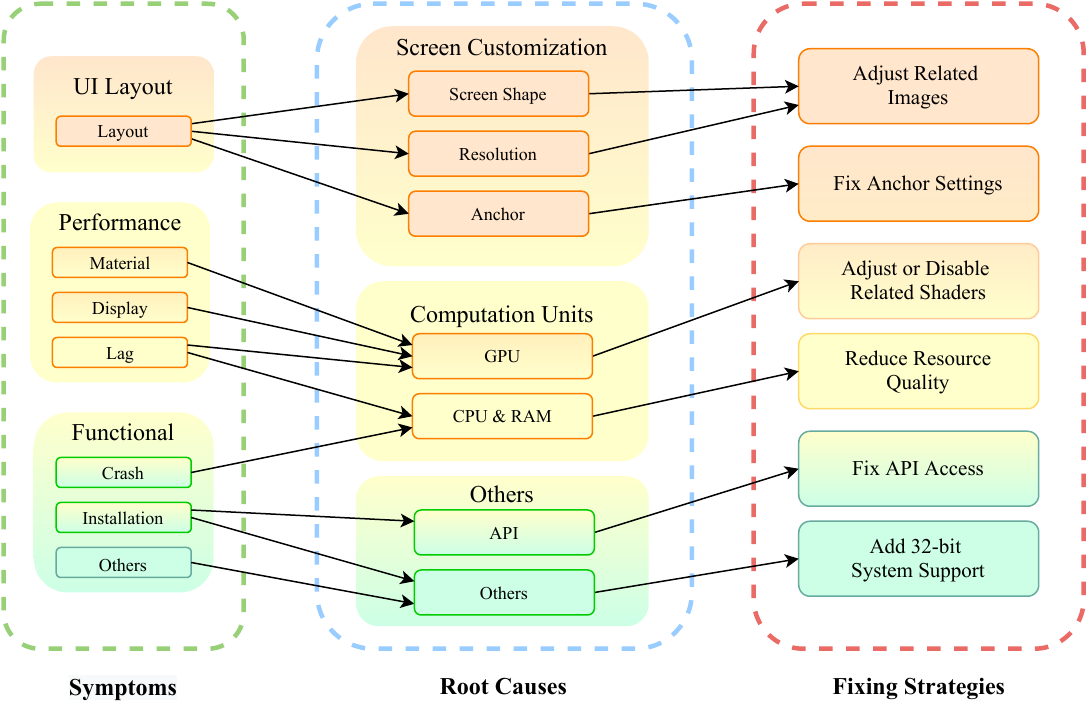}
\caption{A bird view on relationship of issue symptoms, root causes and fixing strategies}
\vspace{-2mm}
\label{fig:map}
\end{figure*}

\subsection{Subject Dataset Collection}
For this study, we reach out and eventually obtain the internal bug reports of four selected games from the testing team of \netease 
First, we discard issues that are duplicated or happening at the early development stage, because during this time slot the game package is not stable and the root causes are hard to be traced. We mainly focus on the bug report in the Alpha test stage, where compatibility tests are generally concentrated. To be noticed, during the game development process, the compatibility testing is not closely associated with the commits, since compatibility tests are performed at certain milestones instead of each regular commit. As a result, our bug reports come from the bug report database instead of the commit report database.
In order to collect effective samples for empirical analysis, we retrieve the bug reports in two different ways. 
We (a) extract all the bug reports with the tag ``compatibility issue'', obtaining about 100 reports, and (b) run a keyword search with keywords related to compatibility issues (\eg``compatibility'', ``corruption'', ``black screen'', \etc), finding over 280 reports. Because the bug report descriptions are not highly standardized, some compatibility issues may not be correctly labeled with the ``compatibility issue'' tag. We put these two retrieved results together for the manual inspection. Two experienced game testers manually check each bug report respectively in order to increase the confidence of the inspection.
As the result, 274 non-duplicated compatibility issues are eventually manually confirmed, forming our study subject dataset of mobile game compatibility issues.
To answer the three research questions posed in Section ~\ref{sec:intro}, for each sample, we analyze the corresponding source code and commits in terms of (a) the symptoms of the issue, (b) the main root cause of the issue, and (c) a brief fixing solution. Because the needed information of some examples is not well recorded in the bug reports, we have to further communicate with the game developers to get sufficient information. Eventually, we end up with 91 distinct compatibility issues with sufficient information that allow us to perform the study. To help future research and development, we open source our dataset used in this paper~\cite{dataset}.

\subsection{Analysis Overview}

For each of the 91 compatibility issues,  our goal is to answer the following set of questions: (a) What is the symptom of the issue? (b) What is the root cause of the compatibility issue? (c) How is the issue fixed according to the root cause? 
Fig.~\ref{fig:map} shows the overview of our findings on these three questions.
The common symptoms of these compatibility issues can be summarized into three categories, UI layout issues, performance issues, and functional issues. Each of the symptom categories is related to one type of root cause and common fixing patterns. For example, most of the performance issues are caused by incompatible GPUs, so the developers could adjust related UI shaders or disable high-level rendering techniques to avoid certain issues. By mapping the relationship between issue symptoms, root causes and common fixing strategies, developers and human testers can quickly locate the root causes and design the fixing strategies for certain symptoms, speeding up the development process. Moreover, we think these findings can help future researchers develop automated testing tools.
The details of each question will be discussed in the following sections.

%% file: manifestation.tex
To facilitate the automated detection of compatibility issues, in this section, we reproduce and analyze the symptoms of the selected compatibility issues and classify them into three main categories: UI layout issues, performance issues, and functional issues. Understanding the common symptoms of mobile game compatibility issues can help developers and manual testers detect and repair such problems. Furthermore, the findings from our analysis can facilitate the development of automatic techniques for compatibility issue detection.  We summarize our categories and distribution in Table~\ref{tab:symptom}. In the rest of this section, we elaborate on each of them in detail.

\begin{tcolorbox}[size=title, colback=white]
{\textbf{Answer to RQ1.} The symptom of game compatibility issues can be categorized into three main categories: UI layout issues, performance issues, and functional issues. The most common symptom of mobile game compatibility issues is UI layout issues, having as many as 53\% of the studied issues (\ie 48 out of 91). Performance and functional issues account for 26\% and 21\% respectively. The distribution of compatibility issues in mobile games is different from general Android software, where functional and performance issues account for 84\% and 4\% of 191 Android compatibility issues, respectively~\cite{wei2016}.
}
\end{tcolorbox}

\subsection{UI Layout Issues}
\label{sec:hardware}
\input{4-hardware.tex}

\subsection{Performance Issues}
\label{sec:perform}
\input{4-perform.tex}

\subsection{Functional Issues}
\label{sec:function}
\input{4-function.tex}

%% file: 4-hardware.tex
Screen layout issues refer to the situation in that UI components are displayed in unusual layouts. This is the most common symptom shown in compatibility issues, we found that 48 out of 91 (53\%) issues belong to this type. In most cases, these compatibility issues appear as tiny layout issues of UI components and do not interrupt the game process. Based on our analysis, these issues can be classified into three subcategories:

\begin{figure*}[bhpt]
\centering
\subfigure[]{
\includegraphics[width=3.7cm]{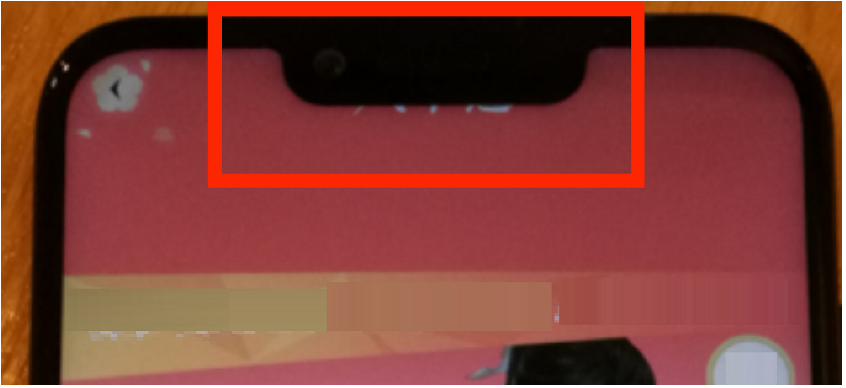}
}
\quad
\subfigure[]{
\includegraphics[width=3.7cm]{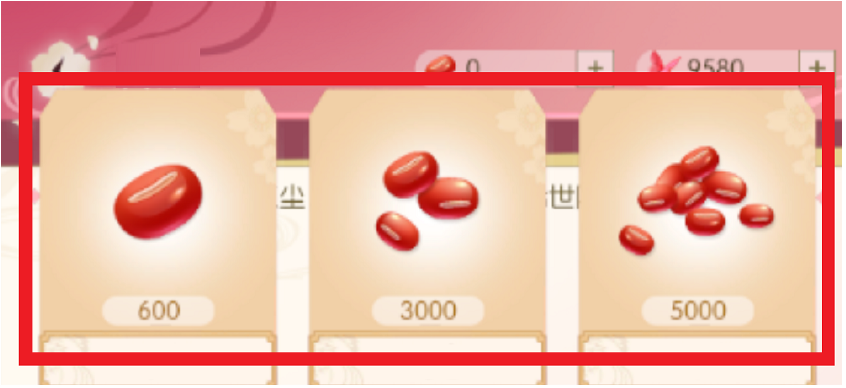}
}
\quad
\subfigure[]{
\includegraphics[width=3.7cm]{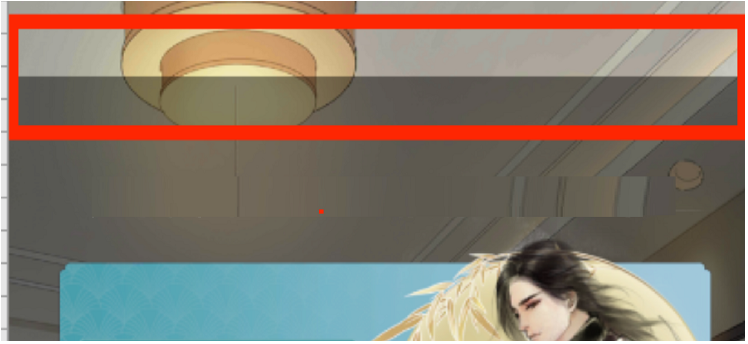}
}
\quad
\subfigure[]{
\includegraphics[width=3.7cm]{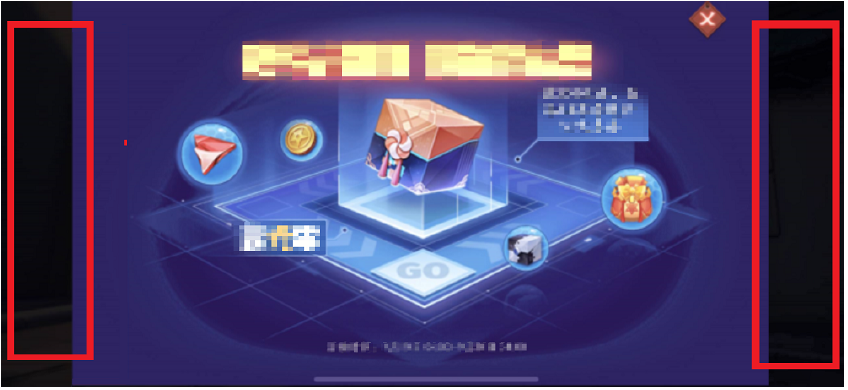}
}
\caption{Examples of screen layout issues. (a) Front camera and speaker are blocking the screen title; (b) UI panel overlaps with an information panel, blocking other buttons; (c) Shadow bar fails to cover the top of screen; (d) Background image fails to cover the whole phone screen, leaving black borders on both sides.}
\label{fig:layout}
\end{figure*}

\subsubsection{Irregular-shaped screen blocking UI components issues}
\ 
\vspace{-4mm}
\newline
\indent Full-screen devices by various manufacturers may have different arrangements of the front camera, which can block some top UI components. Fig.~\ref{fig:layout}~(a) shows an example of such issues. The screen title is fully blocked by the front camera and speaker.


\subsubsection{UI components incorrectly overlapping issues}
\ 
\newline
\indent In some issues, UI components overlap with each other, causing the users unable to obtain information or interact with the blocked buttons in the underlying UI elements. For example, in Fig.~\ref{fig:layout}~(b), the purchase panel is overlapping with the player information interface, so the player cannot touch buttons on the interface. 

\subsubsection{Incomplete occlusion of UI components issues} 
\ 
\newline
\indent Due to the different resolutions and sizes of different devices, UI panel cannot completely cover the expected interface, leaving a blank area on the screen and making the underlying image visible. Fig.~\ref{fig:layout}~(c) shows an example of this type of issue. The grey transparent layer does not completely cover the underlying background image. In other cases, if the background image of the game cannot completely cover the device screen, black borders will appear on one or both sides of the screen. This problem can occur in both horizontal-screen and vertical-screen modes. Fig.~\ref{fig:layout}~(d) shows an example in horizontal mode. 

\vspace{1mm}
\begin{tcolorbox}[size=title, colback=white]
{\textbf{Implication.} For different symptoms, we believe that different automated detection techniques can be applied to UI layout issues: 
(a) Layout issues caused by irregular-shaped screens are usually not directly detectable by image recognition on device screenshots. One straightforward automated solution is to check whether any content will be located in the area of the front camera and speaker.
(b) Currently, there are detection techniques based on image recognition for component occlusion, text overlap and other layout problems existing in general android apps~\cite{nighthawk}. General mobile apps normally have standardized UI layouts, such as upper and lower menu bars, consistent image size and text format, etc. However, the large number of irregular UI components makes it challenging to detect overlapping components in mobile games, especially on 3D game scenes or those games that focus on plot and art style. A potential way is to compare the UI layout of the same gaming scenario on multiple devices~\cite{ki2019}.
(c) For incomplete-occlusion component issues, because it usually has a clear separation line around the screen borders, we suggest that edge detection techniques can be applied to detect this type of issue automatically.
}
\end{tcolorbox}

%% file: 4-perform.tex
The second kind of common symptom appears as performance issues. Unlike UI layout issues, UI exceptions in performance issues are not related to layouts, but display exceptions of a single component or even entire screen, e.g., components disappearing or screen blurred, etc. These issues are mostly related to the GPU performance that can slow down the game process. There are 24 out of 91(26\%) issues in this category. According to different manifestations, performance issues can be divided into three subcategories.

\subsubsection{Material issues}
\begin{figure}[t]
\centering
\subfigure[]{
\includegraphics[width=3.6cm]{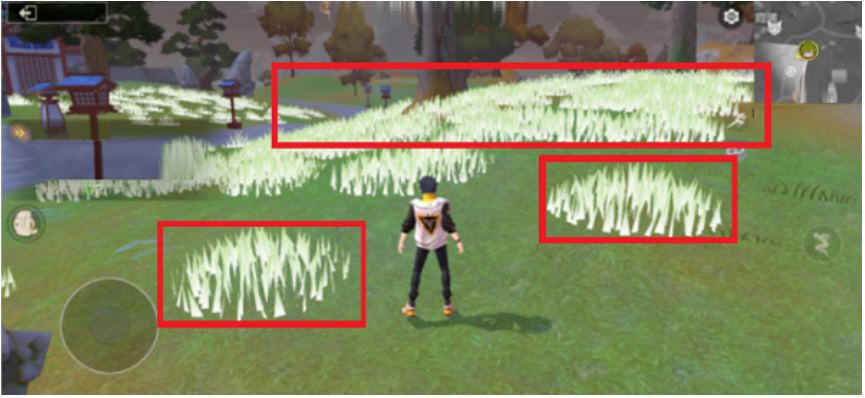}
}
\quad
\subfigure[]{
\includegraphics[width=3.6cm]{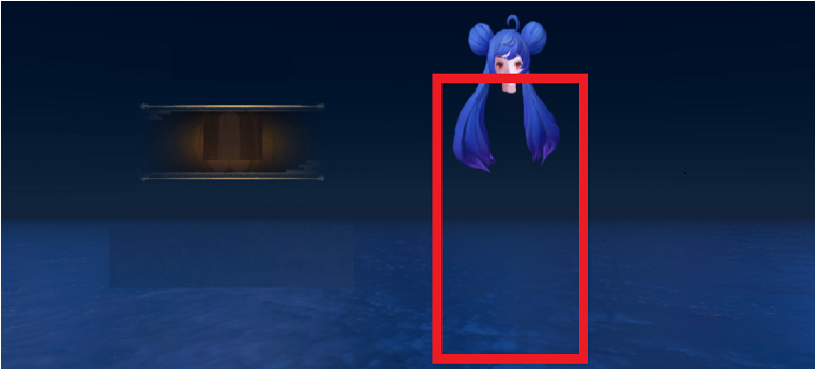}
}
\vspace{-2mm}
\caption{Examples of material issues. (a) Grass material appear as glowing white instead of matte green; (b) Body materials of the character are missing.}
\vspace{-2mm}
\label{fig:material}
\end{figure}

\ 
\newline
\indent 14 out of 24 performance issues belong to this subcategory. Different from the screen adaption issues, the size and position of UI components are correct here, but the material texture of UI components are rendered abnormally, which mainly falls into the following two types:

\begin{itemize}[leftmargin=*]
\item \textbf{Material exception.} 13 of 14 material issues belong to this category. A UI component is displaying on the screen, but its texture does not match the expected performance. As Fig.~\ref{fig:material}~(a), the grass is shown as white glowing texture instead of matte light green.

\item \textbf{Material missing.} Only one of the 14 material issues falls into this category. Since the symptom is inconsistent with others, we separate it out. As the example in Fig.~\ref{fig:material}~(b), the UI material of the character body in the red rectangle disappears on the screen, players can only see the head of the character.

\end{itemize}

\vspace{-3mm}
\begin{table}[htbp]
\centering
\caption{The Distribution of issue symptoms}
\label{tab:symptom}
\begin{tabular}{@{}l|lcc@{}}
\toprule
\textbf{Category} & \textbf{Subcategory}  & \textbf{Count} & \textbf{Total} \\ \hline
\multirow{3}{*}{\begin{tabular}[c]{@{}l@{}}UI\\ Layout\end{tabular}}         & Screen block   & 18  & \multirow{3}{*}{48} \\
    & Overlap  & 16    & \\
   & Incomplete occlusion & 14  &\\ \hline
\multirow{3}{*}{\begin{tabular}[c]{@{}c@{}}Performance\end{tabular}} & Material  & 14  &  \\
& Display    & 5  & 24   \\
& Lag        & 5  &  \\ \hline
\multirow{3}{*}{\begin{tabular}[c]{@{}c@{}}Functional\end{tabular}}  & Startup \& Installation & 9 &\\
& Crash \& ANR  & 5   & 19   \\
& Others & 5 &   \\ \bottomrule
\end{tabular}
\end{table}

\subsubsection{Display issues}
\ 
\newline
\indent Five out of 24 performance issues are display issues. Display issues refer to the circumstances in which the display is abnormal on the whole screen. This is a rarely happened issue, but once occurs, it will immediately interrupt the gameplay and severely harms players' gaming experience. Commonly, since the display issue cannot resume automatically,  players must restart the whole game to restore the display, which is likely to cause the player to lose current game progress. These five display issues have two different manifestations:

\begin{itemize}[leftmargin=*]

\item \textbf{Screen corruption.} Screen corruption shows as all UI components on the screen are displayed incorrectly, including various aspects such as color, position, texture, \etc Fig.~\ref{fig:color}~(a) shows an example of screen corruption. It may occur when: (a) loading into a new game interface, (b) reactivating the game after screen locking, or (c) switching back to the game interface from external pages. We can set specific detecting scenarios for this type of issue based on the happening frequency.

\item \textbf{Screen color exception.} Abnormal screen color includes overexposure screens and gray-scale screens. Fig.~\ref{fig:color}~(b) shows an example of an overexposure screen, the overexposure texture makes players unable to distinguish the position and action of the character. 
\end{itemize}

\subsubsection{Lag issues}
\ 
\newline
\indent Five issues can cause game lag problems. The games can be stuttering or dropping frames, which affects users' game experience, especially for the games that require quick responses, such as ACT game \emph{BS}. 

\vspace{1mm}
\begin{tcolorbox}[size=title, colback=white]
{\textbf{Implication.} Human testers can easily distinguish all three types of performance issues from normal game screens or frames. However, there are no generalized rules for automated testing tools to identify performance issues for all games. 
(a) The current solution to missing images or text in general mobile apps is to train the detector with datasets of screenshots that contain icons representing image broken or the NULL string in the TextView component~\cite{nighthawk}. The material exception or missing issues in games usually don't give obvious signs like icons or strings. Moreover, even with large training samples, learning-based techniques may not work well for detecting performance issues without proper human prior knowledge. Due to different game settings, the issue-looking material in one game scenario may be perfectly normal in other game scenes. For example, the erroneous regions shown in Fig.~\ref{fig:material}~(a) could be normal in other scenarios, e.g. representing interactive materials.
(b) Screen corruption or lagging issues rarely happen in general mobile apps. According to a recent study~\cite{nighthawk}, among the 4,470 screenshots with UI issues, only 1\% of UI issues are related to blurred screen. The current method for general app performance problem detection is to identify unusual battery and heap usage by recording usage on multiple devices~\cite{ki2019}.}

\end{tcolorbox}

%% file: 4-function.tex
Compared with general mobile phone software~\cite{wei2016}, there are fewer functional issues in mobile games.
Among 91 issues, only 19 (21\%) fall into this category. Functional issues refer to the situation that the game fails to install, start, or run as expected, \eg crashing during the gameplay.

\subsubsection{Installation \& Startup issues}
\ 
\newline
\indent Nine out of 19 issues are happening during installation or startup. issues are classified into this type when: (a) the game package fails to install, (b) the installation process does not end even though the package has been installed successfully, (c) the game is successfully installed but cannot open, (d) the game crashes immediately after startup, (e) the game keeps showing a black screen after startup. Same as other applications, this type of issue can be detected easily with testing scripts.

\subsubsection{Crash \& ANR issues}
\ 
\newline
\indent Five of 19 functional issues will cause app crash or application not responding (ANR). According to the behaviors during crashes, crash issues have two types: (a) game crashes with throwing an exception, these issues are easier to catch and repair based on the log, and (b) game suddenly quit and back to the device home screen without any notification. These issues are difficult to catch and figure out the root cause. ANR issue refers to the situation where the game is stuck. If a certain game activity has not been completed in a certain time period, the game will freeze, then an ANR notice will appear on the screen. These issues normally have detailed logs and stack information that can help developers locate and repair bugs.

Crash and ANR issues often appear when players are: (a)  playing or skipping the opening CG after startup, (b) loading into the new game scene, (c) playing in multi-player battle scenes, (d) switching game resolution, (e) selecting and logging in to the game server, (f) downloading and updating in the game.
\begin{figure}[t]
\centering
\subfigure[]{
\includegraphics[width=3.8cm]{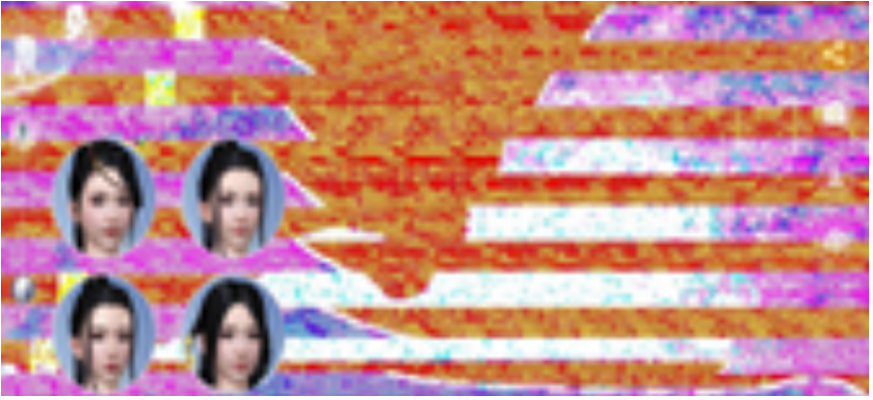}
}
\quad
\subfigure[]{
\includegraphics[width=3.8cm]{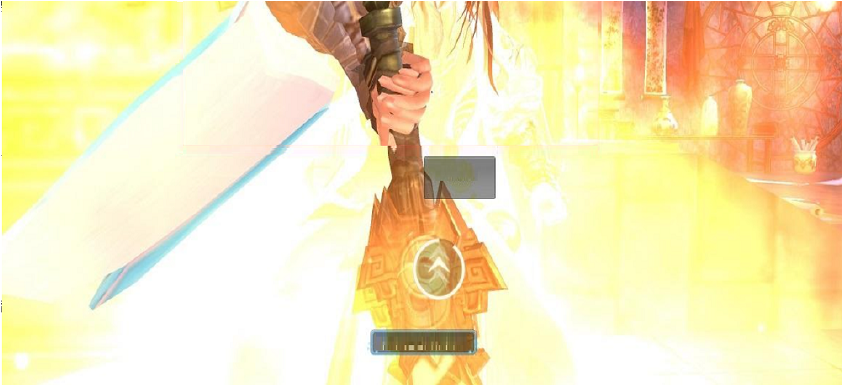}
}
\vspace{-2mm}
\caption{The examples of display issues. (a) Screen corruption; (b) Screen overexposure.}
\vspace{-4mm}
\label{fig:color}
\end{figure}

\begin{tcolorbox}[size=title, colback=white]
{\textbf{Implication.} Similar to general mobile software, functional issues have obvious symptoms, such as installation failure or game crashes, which makes it easy to capture and trace~\cite{wei2016,Cai2019}}. At the same time, functional issues commonly appear in specific scenarios, which is also conducive to applying automatic detection techniques.
\end{tcolorbox}

%% file: rootcause.tex
In this section, we analyze the root causes of these compatibility issues and classify them into three categories. Figuring out the root causes of game compatibility issues can help fix related issues and avoid such problems from happening again.

At first, we followed an approach adopted by Wei \etal~\cite{wei2016}, categorizing the issues into two general
types: \textit{device-specific} and \textit{non-device-specific}. The \textit{device-specific} issues only appear on a certain device model, while \textit{non-device-specific} issues occur on most device models, which are usually caused by a certain API level. Then we further categorized issues in each type into subcategories based on respective root causes. 

However, according to our observation, only three samples in our dataset are \textit{non-device-specific}. In this work, we divided \textit{device-specific} issues into two main categories (Display and Computation Units) and put the \textit{non-device-specific} issues caused by certain systems or API versions into the third category Others.
We summarize categories and distribution in Table~\ref{tab:rootcause}.

\begin{tcolorbox}[size=title, colback=white]
{\textbf{Answer to RQ2.} 
In this study, the root causes of mobile game compatibility issues are basically relevant to the symptoms. As a result, the root causes can also be classified into three main categories: screen customization, computation units, and other reasons including API or system version. Among the 91 issues, 53\% issues (\ie 48) are due to screen customization, and 30\% issues (\ie 27) are caused by computation units. Different from normal Android software, most compatibility issues in mobile games are caused by hardware rather than software.
}
\end{tcolorbox}

\begin{table}[thbp]
\centering
\vspace{-5mm}
\caption{The Distribution of Issue Root Causes}

\label{tab:rootcause}
\begin{tabular}{l|lcc}
\toprule
\textbf{Category}                  & \textbf{Subcategory} & \textbf{Count} & \textbf{Total}      \\ \hline
\multirow{3}{*}{\begin{tabular}[l]{@{}l@{}}Screen\\Customization\end{tabular}}  & Screen Shape & 18 & \multirow{3}{*}{48} \\
                                   & Anchor               & 16             &                     \\
                                   & Resolution           & 14             &                     \\ \hline
\multirow{2}{*}{\begin{tabular}[l]{@{}l@{}}Computation\\Units\end{tabular}} & GPU & 21  & \multirow{2}{*}{27} \\
                                   & CPU \& RAM           & 6              &                     \\ \hline
\multirow{3}{*}{Others}            & API    & 5              & \multirow{3}{*}{16} \\
                                   & Android \& System                 & 3              &                     \\
                                   & Others               & 8              &                     \\ \bottomrule
\end{tabular}
\end{table}
\vspace{-3mm}

\begin{figure}[t]
\centering
\subfigure[]{
\includegraphics[width=3.8cm]{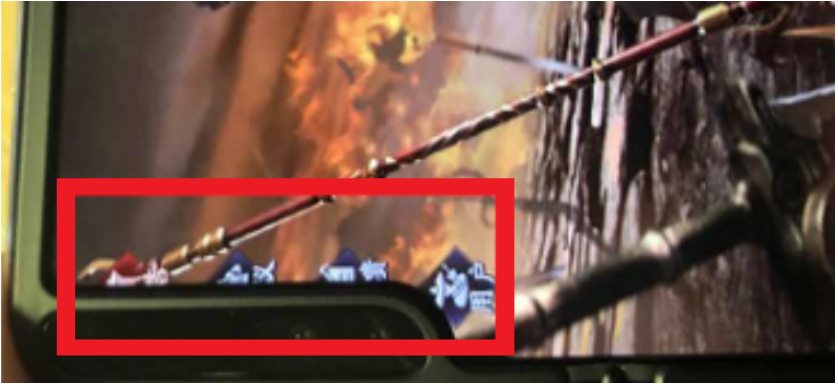}
}
\quad
\subfigure[]{
\includegraphics[width=3.8cm]{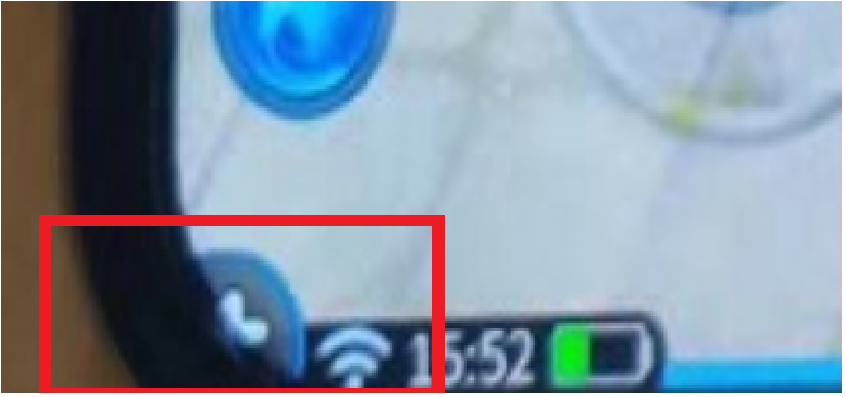}
}

\caption{Examples of display issues. 
(a) Four buttons are blocked by the camera notch; (b) The button of game menu is blocked by the rounded screen corner.}
\vspace{-5mm}
\label{fig:display}
\end{figure}

\subsection{Compatibility of Screen Customization}
\label{sec:display}
\input{3-display.tex}

\subsection{Compatibility of Computation Units}
\label{sec:gpu}
\input{3-gpu.tex}

\subsection{Others}
\label{sec:sdk}
\input{3-sdk.tex}

%% file: 3-display.tex
The majority of compatibility issues are caused by screen customization, we found 48 out of 91 (53\%) compatibility issues belong to this type. This root cause category is exactly corresponding to the screen adaptation issues in Section~\ref{sec:hardware}. These issues can be divided into three subcategories.

\subsubsection{Irregular screen shape}
\ 
\newline
\indent 18 out of 48 issues are in this subcategory, mainly because different manufacturers have launched full-screen mobile phones with different display screen shape designs (\eg the top-notch and bottom home indicator of iPhone full-screen models). These designs could cover the nearby game UI elements, preventing game players to get the game information or press relevant buttons. For example, in Fig.~\ref{fig:display}~(a), four buttons on screen are blocked by the front camera notch on Samsung W20 model, causing the players unable to open menus. Fig.~\ref{fig:display}~(b) shows another example that the game button is blocked by rounded screen corner. 18 issues in this subcategory indicate that developers should consider the irregular screen shape carefully during the development process.

\subsubsection{Incompatible resolution}
\ 
\newline
\indent 16 out of 48 issues occurred due to incompatible image resolution. 
For instance, as we mentioned in Section~\ref{sec:hardware}, the black bars in Fig.~\ref{fig:layout}~(d) appear on both sides of the screen because the background image was prepared for earlier devices instead of full-screen models. This issue occurs quite frequently in recent models.

\subsubsection{Anchor dislocation}
\begin{figure}[thbp]
\centering
\subfigure[]{
\includegraphics[width=2.5cm]{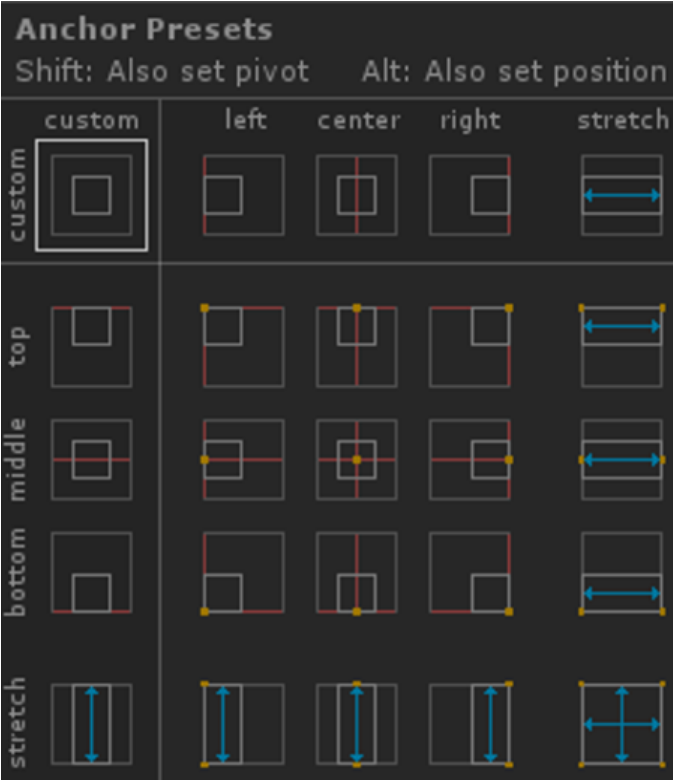}
}
\quad
\subfigure[]{
\includegraphics[width=3.4cm]{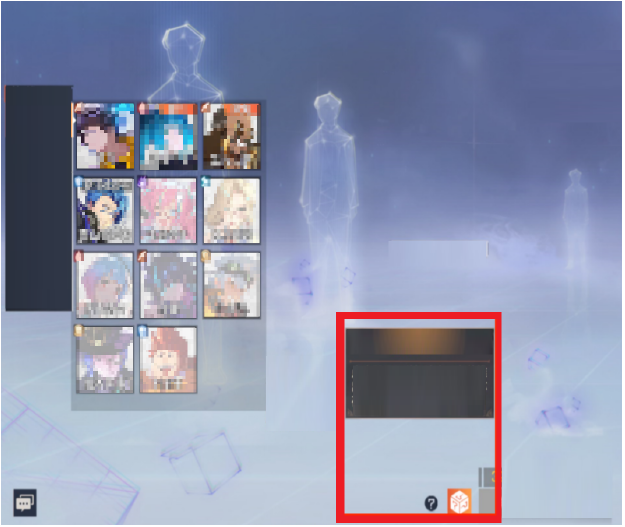}
}
\quad
\subfigure[]{
\includegraphics[width=4cm]{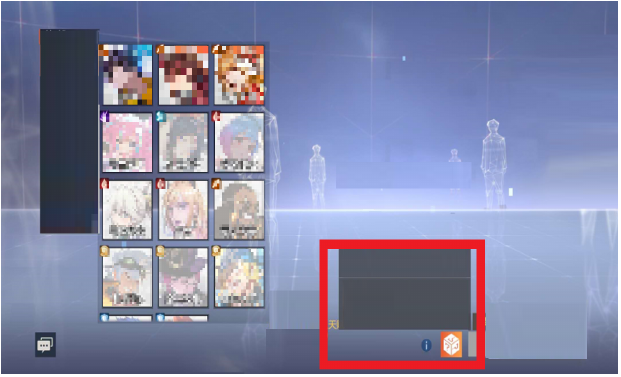}
}

\caption{Examples of anchor dislocation issues. (a) The anchor setting panel in Unity. (b) The black panel in the red rectangle is in the wrong location on the iPad screen due to fixed anchor settings. (c) The correct position for the black panel on a regular resolution screen.}
\vspace{-3mm}
\label{fig:anchorset}
\end{figure}

\ 
\newline
\indent 14 out of 48 issues occur because of Anchor Dislocation. Many game engines (\eg Unity and Cocos), use anchors to set up the position of UI components on the screen. Fig.~\ref{fig:anchorset}~(a) shows the anchor panel in Unity engine. If the anchor of UI element is set in an improper way like hard-codding, the UI element may appear in the wrong position on certain model screens. For example, Fig.~\ref{fig:anchorset}~(b) is a screenshot of an iPad screen, the black panel in the red rectangle should have been located on the right bottom corner, as the same location in Fig.~\ref{fig:anchorset}~(c). The panel was dislocated because its anchor was set by fixed values, which are calculated with a regular-sized screen.

\begin{tcolorbox}[size=title, colback=white]
{\textbf{Implication.} 
Screen customization issues are strongly relevant to the symptom category layout issues. Therefore, once UI layout issues are detected, we can quickly locate and debug them based on the relevant root causes. At the same time, for certain issues caused by anchor dislocation and incompatible resolution, we can perform static analysis techniques to detect and debug them without running the game to save testing costs.
}
\end{tcolorbox}

%% file: 3-gpu.tex
The difference in computation units equipped on the models also has a huge impact on game compatibility testing. Complicated games commonly require high-end device configurations. If the issue is caused by the computation units (\eg CPU, GPU, and RAM) which cannot afford the game performance or be incompatible with game components, we will classify it into this category. As a result, we found 27 (30\%) compatibility issues.

\begin{figure}[bhtp]
\centering
\subfigure[]{
\includegraphics[width=3cm]{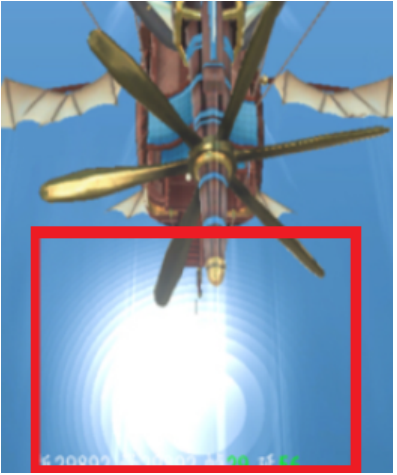}
}
\quad
\subfigure[]{
\includegraphics[width=2.7cm]{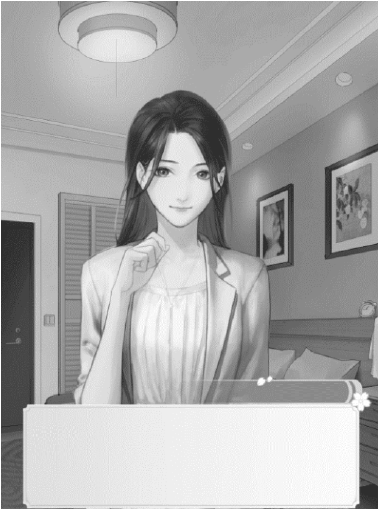}
}
\caption{Examples of GPU performance issue. 
(a) Sunlight in red rectangle exhibits wave texture due to low GPU configuration; (b) Screen shows gray scale due to incompatible shaders.}
\label{fig:gpu}
\end{figure}
\vspace{-3mm}

\subsubsection{GPU issues}
\ 
\newline
\indent 21 of 27 issues in this root cause category are related to GPU. They can be classified into the following two categories:

\begin{itemize}[leftmargin=*]

\item \textbf{Insufficient GPU.} 11 cases belong to this subcategory. This kind of case commonly arises in old models and low-configuration models, they will perform as abnormal game displays, such as in Fig.~\ref{fig:gpu}~(a), the sunlight in the red rectangle has wave texture because the old GPU cannot support high-precision floating numbers. Also, if the game display uses rendering techniques that require high GPU compute capability, many low-configuration models will suffer from game lag. For instance, the grass in Fig.~\ref{fig:material}~(a) uses GPU Instancing technique to enhance realism, this technique can cause obvious lag on low-configuration models, which seriously affects the game experience of players. 

\item \textbf{Incompatible GPU.} In other cases, GPU on the specific model may be incompatible with game settings. For example, the developer of game \emph{LJ} designed unique shaders for the game, but the game will show gray-scale screens on the model VIVO X5M due to different floating-point calculations, as shown in Fig.~\ref{fig:gpu}~(b).
\end{itemize}

\subsubsection{CPU \& RAM issues}
\ 
\newline
\indent In addition to GPU performance issues, CPU and RAM can also affect game compatibility. Six out of 27 issues are related to CPU or RAM. When the game process needs high memory usage, the game may quit unexpectedly on models with small RAM capacity. High CPU overhead will also lead to this problem, like the game scene in Fig.~\ref{fig:cpu}~(a). This scene contains lots of UI components, which is prone to cause crashes on non-flagship and low-configuration models because loading a large number of UI components needs expensive draw calls. Fig.~\ref{fig:cpu}~(b), plotted by Unity Profiler \cite{unitypro}, shows the number of draw calls for loading this scene, which is up to 400.

\begin{tcolorbox}[size=title, colback=white]
{\textbf{Implication.} 
Most performance issues are caused by GPU. For the game components that require high-level configuration, developers are recommended to optimize computation algorithms or providing low-level alternatives. In addition, the mobile games should be extensively tested on broad coverage of various devices with different computational power, so that to identify and address performance issues at an early stage.
}
\end{tcolorbox}

%% file: 3-sdk.tex
16 out of 91 issues are classified into this category. Among them, three are caused by incompatible APIs, and five are related to incompatible mobile systems or Android SDK versions. 
Android ecosystem fragmentation exacerbates this situation. Cases under this category can cause failed installation or crashes during gameplay. We give examples below to describe two main types in this category.

\subsubsection{Incompatible system \& Android SDK version}
\ 
\newline
\indent Incompatible system issues often happen when installing recent games on legacy systems.
Some earlier mobile models still used 32-bit systems, which is not compatible with current 64-bit game installation packages. Games can not be installed successfully under such circumstances. In order to fix it, developers need to recompile the game application, making it compatible with 32-bit systems, which greatly increases the package size. Also, the distribution channel plays an important role in the release stage of mobile games. The game distributors (\eg Google Store), usually have their own development kits, so game developers need to integrate development kits to support account login. However, the API level of development kits may not match mobile games. For example, if the development kit of a certain distributor only supports Android APIs under level 21, an error message will pop up when installing mobile packages with a higher API level.

\subsubsection{Incompatible API}
\ 
\newline
\indent Incompatible API can be a problem on some customized OS. For example, game \emph{LJ} is unable to switch to landscape mode on Google Nexus 6P. Because the related APIs are incompatible.

\begin{tcolorbox}[size=title, colback=white]
{\textbf{Implication.} 
Compatibility issues due to incompatible APIs are very common in general mobile apps. There are several existing works that can detect and fix incompatible APIs through techniques such as static analysis~\cite{mahmud2021, xia2020, zhao2022}, which can also be applied to game testing.}
\end{tcolorbox}

\begin{figure}[bhpt]
\centering
\subfigure[]{
\includegraphics[width=3.8cm]{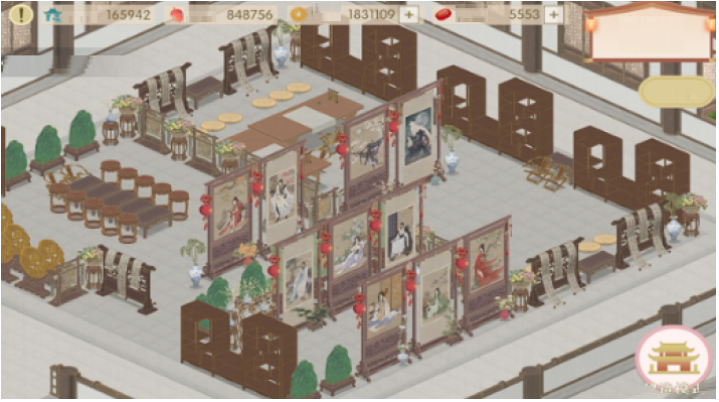}
}
\quad
\subfigure[]{
\includegraphics[width=3.8cm]{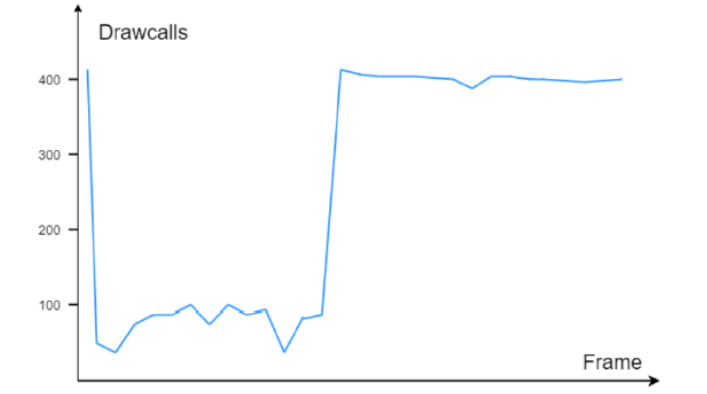}
}

\caption{Example of CPU performance issue. (a) Screenshot of game scenario with a large number of UI components. (b) The number of draw calls when loading the game scene in (a).}
\vspace{-2mm}
\label{fig:cpu}
\end{figure}

%% file: fixing.tex
In this section, we elaborate on the strategies that game developers apply to fix compatibility issues of the various categories.

\subsection{Fixing Screen Customization Issues}
\label{sec:fdisplay}
\input{5-display.tex}

\subsection{Fixing Computation Unit Issues}
\label{sec:fgpu}
\input{5-gpu.tex}

\vspace{-1mm}

\subsection{Others}
\label{sec:fsdk}

\input{5-sdk.tex}

%% file: 5-display.tex
\subsubsection{Irregular screen shape issues}
\ 
\newline
\indent For the layout issues caused by irregular shapes of devices, the developers often adopt the following fixing strategies:

\begin{tcolorbox}[size=title, colback=white]
{\textbf{Answer to RQ3.} 
Besides some API adaption issues, most compatibility issues can be fixed on the developer side. Depending on different root causes, there are various fixing strategies for different issues. Most fixing strategies are still workarounds instead of the complete solutions. Moreover, most of the issues this study involved are fixed manually by developers. We hope that the common fixing patterns of each root cause category can be used as a reference and benefit the further design and development of automated testing techniques for mobile games.
}
\end{tcolorbox}

\textbf{Modify UI layouts based on related models.} 
14 out of 18 irregular shape issues are fixed by adjusting UI layouts. For issues that appear on different models, developers adjust UI layouts based on specific situations. For example, on models with small camera notches (\eg VIVO X21S model) and rounded corners (\eg MEIZU 16th model), developers only adjust the UI components near notches to save effort because the occlusion is minor. In our dataset, 12 related issues are fixed in this way.

On other models with large notches (\eg iPhone) or unique screen ratios (\eg iPad), developers often modify the whole UI layout, moving the buttons and panels on sides closer to the center to ensure notches do not obscure anything.

\textbf{Design screen size SDK.}
To save effort, developers design a Notch SDK to automatically obtain the information of screen notches. The below code snippet is an example message returned by Notch SDK. If the notch (cutout) exists, SDK will send back its size and location.

\begin{lstlisting}[language=json]
"methodId": "getCutoutInfo",
"hasCutout": true,
"width": 300,
"height": 100,
"left": ...
\end{lstlisting}

\textbf{Call API for iPhone home indicator.} 
In particular, Apple Full-screen models since iPhone X replace the home button with a home indicator (\ie a white bar at the bottom of the screen). This home indicator is likely to block information. To avoid blocking game UI, developers will call the related API from game engines to make the home indicator translucent.

\subsubsection{Resolution issues}
\
\newline
\indent For the layout issues caused by incompatible resolution, the developers have two main fixing strategies:

\textbf{Change background image size.} 14 of 16 (88\%) incompatible resolution issues can be fixed by manually modifying the background image size, stretching the image to make it compatible with the irregular-shaped screens, or replacing the background image with a wider one. 

\begin{figure}[hbtp]
\centering
\subfigure[]{
\includegraphics[width=3cm]{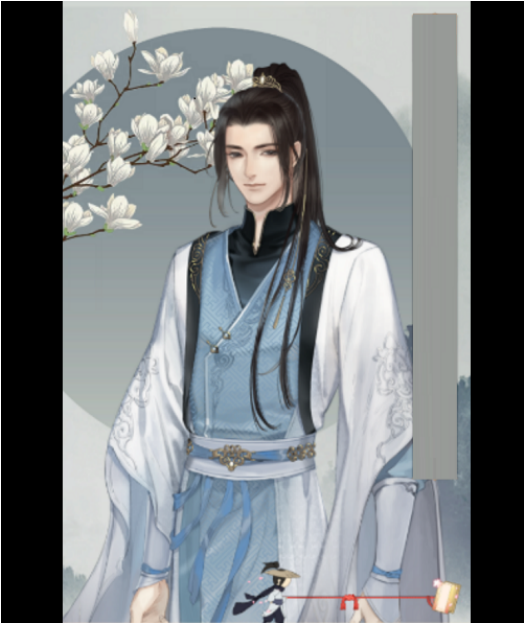}
}
\quad
\subfigure[]{
\includegraphics[width=3cm]{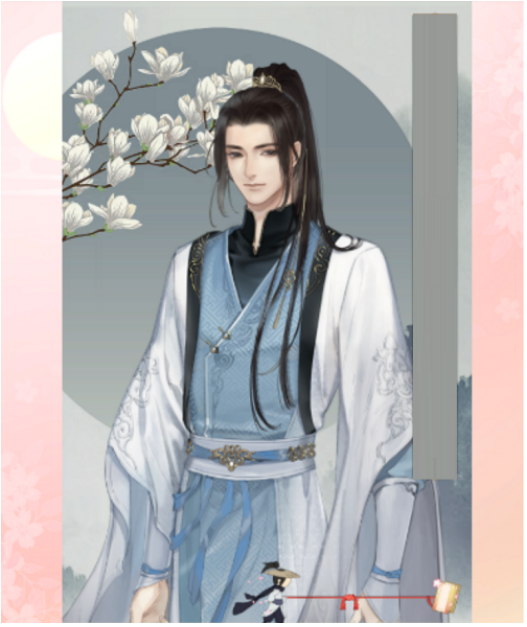}
}

\caption{Example of Fixing resolution issue. (a) Before fixing, the screen has black bars on both side; (b) Adding image in similar style to cover the black bars.}
\vspace{-1mm}
\label{fig:ipad}
\end{figure}

\textbf{Add mask image on the uncovered part.} For circumstances where images cannot be stretched or replaced, developers will add mask images on the uncovered parts to ensure a good appearance of the whole game scene. Fig.~\ref{fig:ipad} shows an example on the iPad version, fixed by adding a wide pink texture under the core image to cover the black background.

\subsubsection{Dislocation issues}
\ 
\newline
\indent \textbf{Modify anchor position.} Adjusting the anchor position can fix all the dislocation issues. During the development phase, in order to avoid anchor dislocation issues, game developers can choose some components which will not be affected by screen sizes, and locate them at first, then use them as references to set other anchors.

\begin{tcolorbox}[size=title, colback=white]
{\textbf{Implication.} 
Device screen SDK could be helpful for the automated game testing tool. With proper knowledge of device screen shape and resolution, layout-related issues could be automatically fixed, e.g., dynamically adjusting background image size, setting the anchor position according to screen shape, etc.
}
\end{tcolorbox}

%% file: 5-gpu.tex
\subsubsection{GPU issues}
\ 
\newline
\indent For the issues caused by GPU performance, the developers basically have the following two fixing strategies:

\textbf{Developing specific shaders for related models.} Adapting low-precision shaders to certain models can fix most UI material issues caused by poor GPU computation performance. For the incompatible GPU, developers choose to modify shaders. For example, as we mentioned in Section~\ref{sec:gpu}, game \emph{LJ} shows gray-scale screens because developers designed unique gray-scale shaders in NGUI (UI kit for Unity), which is in the former code snippet in Fig.~\ref{fig:codegray}. And developers found that the floating-point calculation in this model has a problem that z-axis values of coordinate would be transformed as negative. The NGUI shader will shade the screen in gray-scale when receives negative z-axis values. This issue is fixed by removing the gray-scale shader and each time outputting a gray-scale photo to replace it then, as shown in the latter code snippet in Fig.~\ref{fig:codegray}.
\vspace{-2mm}
\begin{figure}[hbtp]
\centering

\begin{lstlisting}[style=customcsharp]
fixed4 frag (v2f IN) : COLOR{
    half4 col = tex2D(_MainTex, IN, texcoord) * IN.color;
    fixed grey = dot.rgb, fixed3(0.299, 0.587, 0.114);
    col.rgb = lerp(fixed3(grey,grey,grey), col.rgb, step(IN.grey, 0));
    return col;}
\end{lstlisting}

\vspace{1mm}

\begin{lstlisting}[style=customcsharp]
fixed4 frag (v2f IN) : COLOR{
    return tex2D(_MainTex, IN, texcoord) * IN.color;}
\end{lstlisting}

\vspace{-1mm}
\caption{Code snippet of the gray-scale screen problem.}
\label{fig:codegray}
\vspace{-2mm}
\end{figure}

\textbf{Disable rendering techniques.} Disabling those rendering techniques that require high GPU computing performance on low-configuration models can improve lag issues caused by GPU. For example, game lag problem mentioned in section~\ref{sec:gpu} can be solved by disabling GPU Instancing technique from Unity engine.

\subsubsection{Other computation unit issues}
\ 
\newline
\indent To deal with limited CPU overheads and  RAM capacity, developers adopt two strategies on related models.

\textbf{Optimize related algorithms.} Issues caused by limited CPU overheads can be fixed by optimizing related algorithms. For example, the crash issue in Fig.~\ref{fig:cpu}~(a) can be fixed by using Draw Call Batching technique to merge multiple UI components into one draw call. This optimization reduces the number of draw calls, thereby reducing the CPU graphics interface overhead. 

\textbf{Reduce resource quality.} For complicated scenes with many images, reducing the resource quality can fix the insufficient RAM issues. This approach is useful for small memory size models, such as iPhone 6 model with 1 GB Memory.

\begin{tcolorbox}[size=title, colback=white]
{\textbf{Implication.} Providing alternative and multiple candidate game settings under different configurations can be a feasible way to avoid performance issues. Given that the same type of issues on a certain device can appear at different games, testers are recommended to pay more attention to these typical devices.
}
\end{tcolorbox}

%% file: 5-sdk.tex
\textbf{Fixing API access.} For the issues caused by incompatible APIs, developers will choose to adjust related parameters or change the API versions. If some APIs cannot be fixed, developers will add workarounds to keep the main functionalities of game continually working.

\textbf{Add support for 32-bit systems.} Installation issues on 32-bit systems can be fixed by recompiling the installation package and adding support for 32-bit systems. However, this approach will enlarge the installation package. For example, the game package of game \emph{BS} increases 30 MB after adding support for 32-bit systems.

%% file: validity.tex
Our study is empirical and the result is subject to several common threats, including internal, external, and construct validity.

\textbf{Subject Games.} The validity of our empirical analysis may be subject to the threat that our study was only conducted on four mobile games. However, to alleviate this threat, we choose four popular mobile games, all that have been released for more than a year, have large player bases and high daily activity index. The games are also diverse, covering four common types of games (\ie ACT, RPG, MMORPG, and BR)  and containing a bunch of compatibility issues.

\textbf{Selection Criteria.} Our issue selection may also be a potential threat to the validity. As we mentioned in Section~\ref{sec:method}. We collected cases from the bug report database by two strategies, (a) collecting all cases with the tag ``compatibility issue'', and (b) using specific keywords to search through bug reports, such as``compatibility'', ``corruption'', ``black screen'', etc. Also, we narrow down the time frame to only the Alpha test, because the Alpha test is the most intensive time for compatibility testing. We understand that our searching strategy may miss some potential compatibility issues from the database. For example, our keyword search did not include device information, which may cause us to miss some compatibility issues. The reason we did not adopt this strategy is that it is inefficient for our circumstances. Each bug report in the database is suggested to contain the device information. If we use device series as keywords, our result will contain plenty of false positives, greatly increasing the time-consuming of manual inspection. The selection of our search strategies can be justified by their effectiveness and efficiency. Finally, we obtained a total of 274 candidate compatibility issues for further analysis.

\textbf{Manual Inspection.} In order to prevent false positives from our searching results, two experienced game testers from relevant game development teams manually checked each bug report. To be noticed, we followed the widely-used open coding methodology and cross-validated all the cases for consistency.


%% file: related.tex
\subsection{Mobile Compatibility Testing}
Han~\etal~\cite{han12} analyze issue reports from two famous mobile device manufacturers, HTC and Motorola, to study Android compatibility issues. They propose that the main cause of compatibility issues is Android Fragmentation. 
Fazzini~\etal~\cite{fazzini2017} propose the technique that can help developers automatically find cross-platform inconsistencies (CPI) in mobile applications which uses input generation and difference testing to compare software behaviors on different platforms. 
Zhang~\etal~\cite{zhang2019} and Cai~\etal~\cite{Cai2019} carry out empirical research on Android compatibility, respectively. 
Zhang~\etal~\cite{zhang2019} examine the intentions of developers regarding how to achieve software compatibility and avoid potential compatibility issues from benign and malicious apps. 
Cai~\etal~\cite{Cai2019} investigate only benign apps and find that compatibility issues are prevalent and persistent at both installation and run time, most installation incompatibilities are due to API changes during SDK evolution. 
Wei~\etal~\cite{wei2016} analyze 191 fragmentation-induced compatibility~(FIC) issues from real-world open-source Android applications and categorize fragmentation issues according to specific device model, functionalities or performance. They find that most FIC problems are functional problems caused by API incompatibility and propose FicFinder to automatically detect API-related compatibility issues. Huang~\etal~\cite{huang2018} conduct research on Callback Compatibility Issues and subsequently design a static analysis technique Cider to detect related callback compatibility issues. 
In the follow-up work, Huang~\etal~\cite{huang2021} investigate configuration compatibility issues in Android apps and propose ConfDroid analyze XML configuration files to detect compatibility issues automatically. Xia~\etal~\cite{xia2020} propose RAPID to automatically determine whether an API-related compatibility issue has been addressed by combining machine learning with static analysis techniques.
Mahmud~\etal~\cite{mahmud2021} also focus on API-related compatibility issues, they propose ACID to leverage API differences and use static analysis to detect both API invocation and callback compatibility issues.
Liu~\etal~\cite{liu2022} reproduce nine state-of-the-art tools that are proposed to detect compatibility issues in Android apps, they summarize five types of incompatible issues from the results, which indicates that compatibility issues detection is still at an early stage.

\subsection{GUI-related Mobile Compatibility Testing}
The existing works we introduce above are mostly focusing on API-related compatibility issues in Android applications. With the improvement of visual effects in mobile GUI design, UI-related compatibility issues also draw attention from both industry and researchers. Ki~\etal~\cite{ki2019} propose Mimic, an automated UI compatibility testing framework for Android apps. Mimic catches UI anomalies by monitoring differences in the UI hierarchy tree during app running time and calculating the difference between the screenshots by color histogram difference and feature matching. However, Mimic is not suitable for all types of mobile apps, especially for games, because Mimic can not test sensor input. Liu~\etal~\cite{nighthawk} propose NightHawk, an automated tool to detect UI-related compatibility issues, they combine a heuristic-based generation method to automatically generate labeled training data and then use deep learning techniques to model visual information of the GUI screenshot.

\subsection{Mobile Game Testing}
In the past few years, due to the development of reinforcement learning, many pieces of research regarding mobile testing appeared in the world~\cite{feng2020,gu2019icse,david2018,owl2020}, especially in the mobile game aspect\cite{maxim2018,thomas2017}. Lovreto~\etal~\cite{lov2018} randomly choose 16 different mobile games from the list of popular games on Google Play store and analyze mobile game testing by using Appium testing framework and OpenCV library. Khalid~\etal~\cite{khalid2014} conclude player reviews from 99 free mobile games and propose that developers should optimize their limited Quality Assurance (QA) on device models with high priority. Zheng~\etal~\cite{wuji} propose the first game framework that can systematically and automatically test real-world games and obtain great performance. Chen~\etal~\cite{GLIB} propose GLIB, a code-based data augmentation technique, which can effectively uncover GUI non-crash glitches in mobile games.
Even though, it still lacks a comprehensive study on the urgent compatibility issues of mobile game testing. This paper fills this gap and provides important findings that may guide further compatibility testing design for mobile games.

%% file: conclusion.tex
In this paper, we conduct an empirical study to understand and analyze the compatibility issues in mobile games. We investigate 91 issues from 4 popular mobile games to summarize the common symptoms and root causes of compatibility issues, then study the common strategies of fixing issues under different root cause categories. From the empirical study, we obtained some findings and implications that can facilitate the detecting and fixing of compatibility issues in mobile games, hoping to guide and arouse future research for automated compatibility issue detection and fixing.